# Roman CCS White Paper

# The continuous cadence Roman Galactic Bulge survey

**Roman Core Community Survey:** *Galactic Bulge Time Domain Survey*

**Scientific Categories:** *Choose from (descriptions can be found here):*
*exoplanets and exoplanet formation;*
*stellar physics and stellar types;*
*stellar populations and the interstellar medium;*

**Additional scientific keywords:** *Suggestions for each of the above categories can be found*
*Binary stars / Trinary stars, White dwarf stars, Stellar evolution*

**Submitting Author:**
Name: Thomas Kupfer
Affiliation: Texas Tech University
Email: tkupfer@ttu.edu

**List of contributing authors** (including affiliation and email):
- Camilla Danielski, Instituto de Astrofísica de Andalucía (CSIC) - cdanielski@iaa.es
- Poshak Gandhi, University of Southampton - poshak.gandhi@soton.ac.uk
- Tom Maccarone, Texas Tech University - Thomas.Maccarone@ttu.edu
- Gijs Nelemans, Radboud University Nijmegen - nelemans@astro.ru.nl
- Valeriya Korol, Max Planck Institute for Astrophysics - korol@mpa-garching.mpg.de
- Liliana Rivera Sandoval, University of Texas Rio Grande Valley (UTRGV) - liliana.riverasandoval@utrgv.edu

**Abstract:**
Galactic binaries with orbital periods less than 1 hour are strong gravitational wave sources in the mHz regime, ideal for the Laser Interferometer Space Antenna (LISA). At least several hundred, maybe up to a thousand of those binaries are predicted to be sufficiently bright in electromagnetic wavebands to allow detection in both the electromagnetic and the gravitational bands allowing us to perform multi-messenger studies on a statistically significant sample. Theory predicts that a large number of these sources will be located in the Galactic Plane and in particular towards the Galactic Bulge region. Some of these tight binaries may host sub-stellar tertiaries. In this white paper we propose an observing strategy for the Galactic Bulge Time Domain Survey which would use the unique observing capabilities of the Nancy Grace Roman Space telescope to discover and study several 10s of new strong LISA gravitational sources as well as exoplanet candidates around compact white dwarf binaries and other short period variables such as flaring stars, compact pulsators and rotators.

## Introduction

Compact binaries are a class of binary systems with orbital periods below a few hours, consisting of a black hole, neutron star or white dwarf primary and a helium star, white dwarf, or neutron star secondary. The study of these systems is important to our understanding of such diverse areas as supernova Ia progenitors crucial for cosmology and binary evolution. These systems are also predicted to be the dominant gravitational wave sources detectable by the Laser Interferometer Space Antenna (LISA), an approved ESA/NASA mission on track for mission adoption early 2024 (Amaro-Seoane et al. 2022). At least several hundred, maybe even over a thousand of those binaries are predicted to be sufficiently bright in electromagnetic wavebands to allow detection in both the electromagnetic (EM) and the gravitational (GW) bands, providing a unique opportunity to perform multi-messenger studies on a statistically significant sample (Korol et al. 2017, Lamberts et al. 2019). This makes these binaries ideal multi-messenger sources where EM and GW can complement each other and, in some cases, are even required to break degeneracies in the system parameters. The GW signal alone has a strong degeneracy between its amplitude and the system's inclination angle as well as in some cases a poor sky localization of several square degrees and between chirp masses and distance which may affect the low-frequency part of the LISA sample severely. This degeneracy can be solved with prior information on the period, sky localization, mass, distance, and inclination from EM measurements (Shah et al, 2012, 2014; Finch et al. 2023). Thus, multi-messenger observations of these sources have the potential to yield more robust masses, radii, orbital separations, and inclination angles than can be achieved using either GW or EM observations alone. However, the improvement in system properties from multi-messenger studies depends strongly on the signal-to-noise in the LISA data as well as on the precision of the derived EM parameters and their correlations.

Currently, we know only about two dozen of these sources (Kupfer et al. 2023). Their light curves show variations on timescales of the orbital period, e.g., due to eclipses or tidal deformation of the components. Therefore, photometric time-domain surveys are well suited to identify LISA binaries in a homogeneous way. Binary population studies have shown that a large number of these sources are expected to reside in the Galactic Bulge region, ideally located for the Galactic Bulge community survey (Lamberts et al. 2019). The Roman telescope combines unprecedented sky resolution with large photometric depths unreachable from the ground.

## Survey proposal

We propose to use these unique capabilities as part of the Galactic Bulge community survey to conduct a several hours continuous minute cadence survey mainly targeted to discover rapid variability coming from compact gravitational wave sources which are ideal for multi-messenger studies. As such this dataset will provide a unique resource to search for electromagnetic counterparts of binaries discovered by LISA. Additionally, this survey will discover a large sample of compact supernova Ia progenitors and other binary systems with orbital periods below a few hours, including cataclysmic variables, X-ray binaries or compact white dwarf binaries with low mass companions (e.g., Brown et al. 2022, Chen et al. 2020, Kupfer et al. 2020, Burdge et al. 2019). By exploiting the infrared (IR) capabilities of the Roman Telescope we can also study the companions of compact binaries. For example, recent observations of compact white dwarf binaries have revealed IR excesses likely due to their companion stars (Green et al. 2020, Rivera

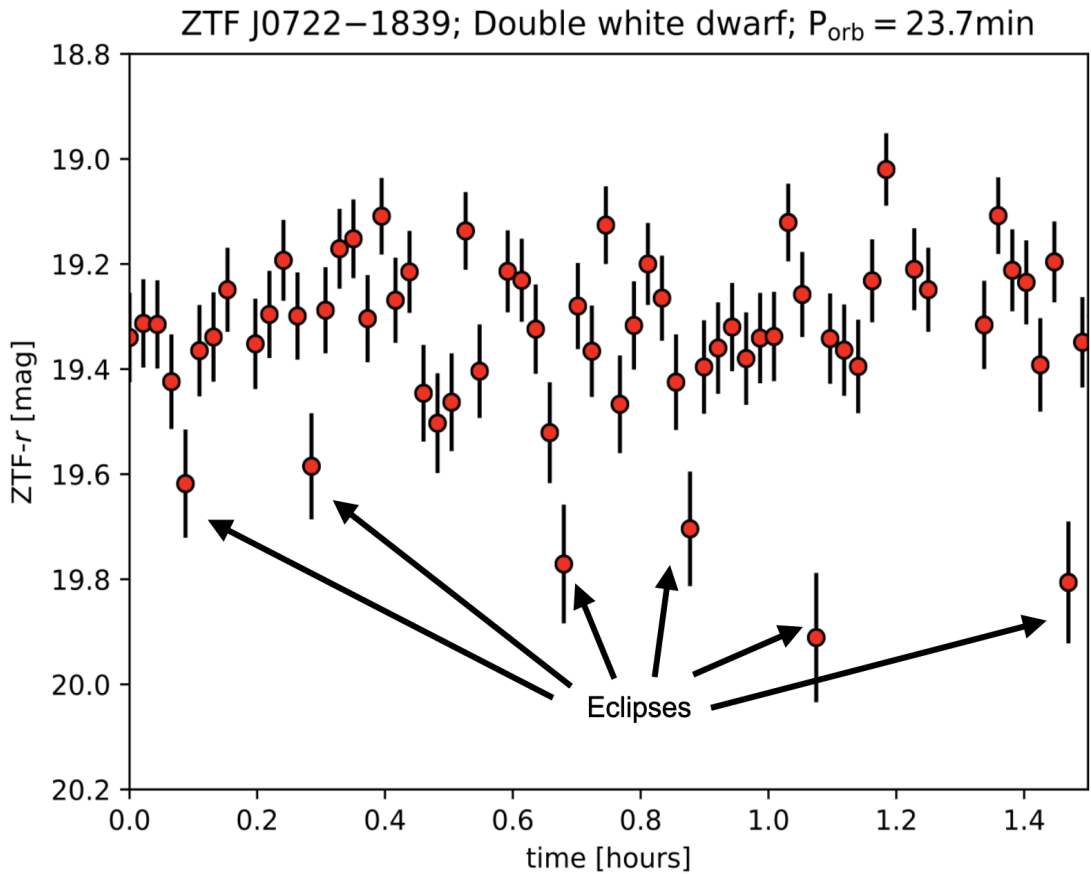

*Figure 1: Continuous minute cadence light curve of a LISA detectable source obtained with the Zwicky Transient Facility. The short eclipse duration emphasizes the importance of continuous minute cadence observations.*

Sandoval et al. 2021, van Roestel et al. 2021). The infrared light emitted by the cold companions is less affected by interstellar extinction and can penetrate the circumstellar material, allowing variability detection towards regions of the Galactic Bulge with high extinction. Infrared light curves can also help to investigate heating on the companions of spider pulsars such as black widows (Draghis et al. 2019). Additionally, we expect to find 1000s of other sources with rapid variability such as flaring stars, compact pulsators, and rapid rotators (e.g., Kupfer et al. 2021). As such, this survey will provide a unique resource to study minute timescale variability in stellar sources. The Roman telescope is the ideal resource to perform such a survey. The sky resolution is at least an order of magnitude better than the seeing-limited ground-based observations, leading to significantly less blending. A 60 sec exposure on Roman will reach a limiting magnitude of 25 mag, providing a large, expected sample of compact LISA binaries, supernova Ia progenitors, flaring stars, compact pulsators and any other object with short period variations between Earth and the Galactic Bulge. Continuous cadence is key to rapid events or events with short duty cycles. Eclipsing LISA binaries with orbital periods less than a few hours have duty cycles of approximately 10% i.e., eclipse durations can be as short as one minute up to a few minutes (see Fig. 1 for an example, taken from Kupfer et al. 2021). Therefore, the currently planned 15 min cadence is not ideal to detect the short period variability of these binaries. We request to use a continuous 60 sec cadence or at least shorten the 15 min cadence to less than a few minutes. We propose to use the Wide Field Instrument and alternate between the F062 and the F087 or any redder filters which would provide additional color information. The F062 and the F087 or any redder filter is the ideal compromise to be sensitive to the hotter white dwarfs and cool companion stars as well as extinction towards the Galactic Bulge region. Based on a study by Digman et al. (2022) and other binary population studies, we expect to find several 10s of LISA binaries.

Furthermore, some of these short period binaries may host sub-stellar tertiaries. As such, our survey will complement the default 15 min cadence Roman Galactic Bulge Time Domain Survey for what it concerns the study of exoplanets and their demography (Penny et al., 2019). Thanks to the magnitude limit of 27 mag (in W146), Roman is, with the regular 15 min cadence, able to search for lensing events from giant exoplanets orbiting compact binary stars. However, such a lensing event would appear as one body lens with a planetary companion, for such we cannot exclude that the host white dwarf star is not an unresolved compact double white dwarf. As such, the continuous minute cadence survey will ideally complement the regular 15 min cadence to identify lensing events from substellar companions around compact binary stars. We know that the probability for planets to survive the binary evolution is higher than in a single-star case, and survival rates around binaries are estimated to be between 20-30% (Columba et al., 2023). For those planets detected by the Galactic Bulge Time Domain Survey within 1 and ~ 3 kpc (observed in the same field/s of our high cadence survey) we will be able to provide constraints on the hosting white dwarfs and for such properly characterize the planets that have been identified by the Galactic Bulge Time Domain Survey. This shows that the minute cadence survey ideally complements the regular 15 min cadence. Detecting planets around compact binaries will provide the first EM observational constraints on the final moment of the life of exoplanets orbiting binary stars, filling up the exoplanet demography science case across the Hertzsprung Russell diagram, and helping to further develop the science of exoplanets with LISA (Danielski et al., 2019).

## **Summary**

Here we propose minute cadence observations with the Nancy Grace Roman Space telescope to study stellar variability on minute timescales as part of the Roman Core Community Survey. This unique cadence will provide a unique resource and is expected to discover flaring stars, compact pulsators, compact rotators, different types of compact binaries as well as a sample of supernova Ia progenitors and several 10s of gravitational wave sources detectable for LISA. We would like to emphasize that continuous cadence observations will provide a unique observing mode for the Roman telescope opening a new window to study rapid variability in dense Galactic Bulge regions with space-based quality and ideally complements the regular 15 min cadence observations. The combination of cadence with diffraction limited observations in the Galactic Bulge probes a very different parameter space than any other survey including the Legacy Survey of Space and Time (LSST).

# References

Amaro-Seoane, P., Andrews, J., Arca Sedda, M., et al. 2022, arXiv e-prints, arXiv:2203.06016
Brown, W. R., Kilic, M., Kosakowski, A., & Gianninas, A. 2022, ApJ, 933, 94
Burdge, K. B., Coughlin, M. W., Fuller, J., et al. 2019, Nature, 571, 528
Chen, W.-C., Liu, D.-D., & Wang, B. 2020, ApJL, 900, L8
Columba, G., Danielski, C., Dorozsmai, A., et al. 2023, arXiv e-prints, arXiv:2305.07057
Danielski, C., Korol, V., Tamanini, N., & Rossi, E. M. 2019, A&A, 632
Digman M. C., Hirata C. M., 2022, arXiv e-prints, arXiv:2212.14887
Draghis, P., Romani, R. W., Filippenko, A. V., et al. 2019, ApJ, 883, 108
Finch, E., Bartolucci, G., Chucherko, D., et al. 2023, MNRAS, 522, 5358
Green, M. J., Marsh, T. R., Carter, P. J., et al. 2020, MNRAS, 496, 1243
Korol, V., Rossi, E. M., Groot, P. J., et al. 2017, MNRAS, 470, 1894
Kupfer, T., Korol, V., Littenberg, T., et al., 2023, arXiv e-prints, arXiv:2302.12719
Kupfer, T., Prince, T. A., van Roestel, J., et al. 2021, MNRAS, 505, 1254
Kupfer, T., Bauer, E. B., Marsh, T. R., et al. 2020, ApJ, 891, 45
Lamberts, A., Blunt, S., Littenberg, T. B., et al. 2019, MNRAS, 490, 5888
Penny, M. T., Gaudi, B. S., Kerins, E., et al. 2019, ApJS, 241, 3
Rivera Sandoval, L. E., Maccarone, T. J., Cavecchi, Y., et al. 2021, MNRAS, 505, 215
Shah, S., van der Sluys, M., & Nelemans, G. 2012, A&A, 544
Shah, S., & Nelemans, G. 2014, ApJ, 790, 161
van Roestel, J., Kupfer, T., Green, M. J., et al. 2021, MNRAS